\documentclass[a4paper]{jpconf}

\usepackage{color,latexsym,graphics,graphicx,epsfig}
\usepackage{graphicx,graphics,color,latexsym}
\usepackage{hyperref}
\usepackage{ifthen}
\usepackage{amsmath}
\usepackage{footmisc}

\def\t{\tilde }
\def\pu{\t p\cdot u}
\def\pr{{\mathcal P}}

\def\nc{{\, ,}}
\def\np{{\, .}}

\def\p{{\bf p}}

\def\st{\begin{equation}}
\def\stp{\end{equation}}
\def\bg{\begin{eqnarray}}
\def\nd{\end{eqnarray}}
\def\Eq#1{Eq.~(\ref{#1})}

\def\Fig#1{Fig.~\ref{#1}}

\def\pr{\mathcal P}
\def\bra{\langle}
\def\ket{\rangle}

\def\llangle{\left\langle}
\def\rrangle{\right\rangle}

\def\Eq#1{Eq.~(\ref{#1})}

\def\Fig#1{Fig.~\ref{#1}}

\begin{document}
\title{Anisotropic Flow and Viscous Hydrodynamics}
\renewcommand\footnotemark{}
\renewcommand\footnoterule{}
\author{Li Yan\footnote{
In collaboration with Derek Teaney.}
}

\address{
Department of Physics and Astronomy, 
Stony Brook University, 
Stony Brook, NY, 11794, USA}

\ead{li.yan@stonybrook.edu}

\begin{abstract}
We report part of our recent work on viscous hydrodynamics with 
consistent phase space distribution $f(x,\p)$ for freeze out. We develop the 
gradient expansion formalism based on kinetic theory, 
and with the constraints from the comparison 
between hydrodynamics and kinetic theory, viscous corrections to $f(x,\p)$ can 
be consistently determined order by order. Then with the obtained $f(x,\p)$, 
second order viscous hydrodynamical calculations are carried out for elliptic flow $v_2$. 
\end{abstract}

\section{Introduction}
Relativistic hydrodynamics is an important theoretical tool in heavy ion collisions. 
Especially it successfully reproduces the observed 
elliptic flow $v_2$. In the past several years, with respect to the updated
understanding of collective phenomena in heavy-ion collisions, 
two major progresses have been made in 
hydrodynamical models. One is the realization of fluctuations in initial 
state\cite{Alver:2010gr}, which leads to the practically used event-by-event hydrodynamics\cite{Schenke:2010rr,Qiu:2011iv}. And
correspondingly many new types of anisotropic flow, such as dipole flow $v_1$\cite{Teaney:2010vd} and 
triangular flow $v_3$\cite{Alver:2010gr,Alver:2010dn}, are  
studied. The other one is the inclusion of viscous corrections\cite{Teaney:2003kp}.

In practice,
although ideal hydrodynamics has been found effective in characterizing the 
collective medium expansion, viscous corrections are not negligible\cite{Danielewicz:1984ww}. 
One particular example
of viscous corrections to hydrodynamics is the viscous damping of elliptic 
flow\cite{Dusling:2007gi,Luzum:2008cw,Song:2007ux}.          

Many attempts have been made in the investigation of the viscous corrections to 
hydrodynamics\cite{Dusling:2007gi,Luzum:2008cw,Song:2007ux}. 
Since first order viscous hydrodynamics suffers from the causality problem(see for example\cite{Romatschke:2009im}), 
most of the established viscous hydrodynamical models have contained second order viscous 
terms in the equations of motion, such as Israel-Stewart theory. Recently, in an important
paper, Baier and his collaborators(BRSSS) developed second order viscous hydrodynamical theory
with conformal symmetry assumption\cite{Baier:2007ix}. Hydrodynamical simulation to the medium expansion
in heavy ion collisions is achieved by solving hydrodynamical equations of motion, with
respect to the specified initial conditions. Observables are obtained then by freeze out 
process. Due to the correspondence between hydrodynamics and kinetic theory, freeze out
process also needs viscous corrections, and the corrections have to be constrained so that
they are consistent with hydrodynamics itself. In particular, in Cooper-Fryer formula, this 
is reflected in the determination of the phase-space distribution function $f(x,\p)$. 
However, to the knowledge of authors, this
constraints have not been systematically discussed in the literature through
second order, and this will be the main
subject of this paper. For more details, refer to \cite{Yan:2012b}. 

The paper is organized as follows. The theoretical formalism is constructed 
by reviewing viscous effects in kinetic theory and hydrodynamics
in section \ref{kt_and_hydro}. In section \ref{df_solution} the  
determination of phase-space distribution function is discussed. 
Hydrodynamical simulations with consistent form of $f(x,\p)$ is 
carried out, and elliptic flow $v_2$ is calculated at RHIC energy, 
in section \ref{hydro_res}. Our conclusions are summarized in section \ref{con}.

\section{Kinetic Theory and Hydrodynamics}
\label{kt_and_hydro}

A liquid system out of equilibrium can be approached through either transport theory 
or hydrodynamics, provided that the non-equilibrium part can be seen as perturbations.
As a result, in order to sustain the continuity around freeze out, there must be a determined 
correspondence between kinetic theory and hydrodynamics. This correspondence originates the 
constraints on distribution function $f(x,\p)$. In terms of the dependence on transverse 
momentum $p_T$, one can generally fix $\delta f$ by the 
moment method\cite{DeGroot:1980dk}, in which 
$\delta f$ is decomposed into moment expansion and cut at a certain 
order. The widely accepted form, for example,
\st
\label{dfpi}
\delta f_\pi=\frac{n_p(1\pm n_p)}{2(e+\pr)T^2}p^\mu p^\nu\pi_{\mu\nu},
\stp
is obtained by a fourteen-momentum method. However there are two flaws in this 
form for realistic simulations. First, in actual simulations $\pi_{\mu\nu}$
is taken from second order hydrodynamical calculations, while 
$\delta f_\pi$ is computed using a first order approximation.  
Second, and more importantly, the $p_T$ dependence in $\delta f$ 
has been investigated in \cite{Dusling:2009df}, and 
found to vary based on microscopic dynamics. Although the quadratic ansatz (i.e. \Eq{dfpi})
is appropriate for collisional energy loss, it fails in the case
where radiative energy loss dominates. 

In kinetic theory, the viscous corrections
$\delta f(x,\p)$ corresponds to the viscous corrections to hydrodynamics, i.e., 
stress tensor $\pi^{\mu\nu}$, so it also depends on gradient expansion. 
In this way, following Chapman-Enskog method\cite{DeGroot:1980dk} where 
expansion in gradients is used to find asymptotic form of the solution to
Boltzmann equation,  $\delta f(x,\p)$ can be determined order by order. 
When relaxation time approximation can be taken into
account, form of $\delta f$ is analytically solvable for the specified viscous hydrodynamics. 
In this paper, we take relaxation time approximation and 
determine $\delta f(x,\p)$ through second order in gradients, with respect to BRSSS
hydrodynamics.

The matrix convention we use throughout the paper is $(-,+,+,+)$. 
$p^\mu$ is used for four-momentum. And for the 
sake of convenience, we introduce a tilde on a quantity from time to time to
indicate it as dimensionless. In the derivation of hydrodynamics, tensor index
is always split into temporal and spatial parts. Since flow four-velocity 
$u^\mu=\gamma(1,\vec v)$ is purely temporal in the local rest frame(LRF), the 
decomposition is then realized based on $u^\mu$ and the projection operator
$\Delta^{\mu\nu}=u^\nu u^\nu+g^{\mu\nu}$. For derivative operator we have 
$\partial^\mu=u^\mu D+\nabla^\mu$, with
$D=u^\mu\partial_\mu$ and $\nabla^\mu=\Delta^{\mu\nu}\partial_\nu$
representing pure time and space derivatives in LRF respectively. 
A further decomposition according
to rotation group is also considered comparing to the existed tensor form in 
hydrodynamics, so the indices of a tensor in brackets $\llangle\ldots\rrangle$ 
stand for being symmetric, (projection operator) traceless and transverse,
\begin{align}
&A^{\mu\ldots\llangle\ldots\alpha\ldots\beta\ldots\rrangle\ldots\nu}
=A^{\mu\ldots\llangle\ldots\beta\ldots\alpha\ldots\rrangle\ldots\nu}\\
&A^{\mu\ldots\llangle\ldots\alpha\ldots\beta\ldots\rrangle\ldots\nu}
\Delta_{\alpha\beta}=0\\
&A^{\mu\ldots\llangle\ldots\alpha\ldots\rrangle\ldots\nu}
\Delta_{\alpha\beta}
=0
\end{align}
More details of the decomposition can be found in \cite{Yan:2012b}. 

\subsection{Hydrodynamics}

In Chapman-Enskog expansion, the so-called solubility condition\cite{DeGroot:1980dk} relates
spatial gradients and temporal derivatives. This is actually equivalent to
hydrodynamical equations of motion $\partial_\mu T^{\mu\nu}=0$. 
In the current section we will derive 
the viscous corrections through second order to BRSSS hydrodynamical equations of
motion. 

The energy-momentum tensor of viscous hydrodynamics reads
\st
\label{Tideal}
  T^{\mu\nu}  = e u^{\mu} u^{\nu} + \pr \Delta^{\mu\nu} + \pi^{\mu\nu} \nc
\stp
where $\pi^{\mu\nu}$ is the viscous correction to stress tensor. 
In this paper, the bulk viscosity
$\zeta=0$ and baryon chemical potential $\mu_B=0$. 
Up to second order viscous corrections, BRSSS\cite{Policastro:2001yc} determined that 
the possible forms of the gradient expansion 
in the stress tensor in a conformal liquid are 
\begin{multline}
\label{stress_2}
\pi^{\mu\nu} = -\eta \sigma^{\mu\nu} 
 + \eta \tau_\pi \left[ \llangle D\sigma^{\mu\nu}\rrangle + \frac{1}{d-1} \sigma^{\mu\nu} \nabla \cdot u \right] 
   +  \lambda_1 
\llangle \sigma^{\mu}_{\phantom{\mu} \lambda} \sigma^{\nu \lambda } \rrangle +  
\lambda_2  \llangle \sigma^\mu_{\phantom{\mu} \lambda} \Omega^{\nu \lambda} \rrangle + 
\lambda_3 \llangle \Omega^\mu_{\phantom{\mu} \lambda} \Omega^{\nu \lambda} \rrangle  \nc
\end{multline}
where $\sigma^{\mu\nu}=2\bra\nabla^\mu u^\nu\ket$ and the vorticity tensor is defined as
\st
 \Omega^{\mu\nu} = \frac{1}{2} \Delta^{\mu\alpha} \Delta^{\nu\beta} \left( \nabla_\alpha u_\beta - \nabla_\beta u_\alpha \right)  \np
\stp
$(\tau_\pi,\lambda_1,\lambda_2,\lambda_3)$ are corresponding second order
transport coefficients.
$d=4$ is the number of space-time dimensions.\footnote{
In this paper, we will ignore the possible effects of higher space-time dimensions,
so we write explicitly wherever $d=4$. Also we ignore all curved space-time 
effects discussed in \cite{Policastro:2001yc}.}
Then equations of motion according to \Eq{stress_2} can be formulated as
\begin{align}
\label{eom}
De =&  -(e + \pr) \nabla \cdot u  + \frac{\eta}{4} \sigma_{\mu\nu} 
\sigma^{\mu\nu} + O(\nabla^3)\nc\\
Du_{\mu} =& -  \frac{\nabla_{\alpha} \pr}{e + \pr}   
   + \frac{u_\mu D\sigma^{\mu \nu} \Delta_{\alpha\nu }}{e+\pr}-
   \frac{\Delta_{\alpha\nu}\nabla_\mu\sigma^{\mu\nu}}{e+\pr} + O(\nabla^3) \np
\end{align}
Note that the temporal derivatives of the flow velocity $u^\mu$ and the energy density 
$e$ are written in terms of spatial gradients through second order, while higher order
terms are neglected.

\subsection{Kinetics}

To determine the viscous corrections to the distribution function, the strategy is to 
solve the kinetic equations in a relaxation time approximation order by order
in the gradient expansion: 
\st
\label{Expansion}
  f(x,\p) \equiv n_\p(x,\p) + \delta f(x,\p) = n_\p(x)  + \delta f_{(1) }  + \delta f_{(2)} +
O(\nabla^3)\nc 
\stp
where $n_\p(x)$ is equilibrium distribution, which depends on the space-time
coordinates through the dimensionless combination $p\cdot u/T=\pu$. 
In classical limit, 
$
 n_\p(x) = \mbox{exp}[p\cdot u(x)/T(x) ].
$
In a relaxation time approximation the Boltzmann equation reads
\st
\label{boltzmann}
 p^{\mu} \partial_{\mu} f_\p(x)  =     -  \frac{T^2}{\t C_\p} \left[ f - n(p\cdot u_{*}(x)/T_{*}(x)) \right]\nc
\stp
where the dimensionless coefficient $\t C_\p$ is related to the canonical momentum dependent relaxation time
$\tau_R$,
\st
 \t C_{\p} = -T^2\,  \frac{\tau_R(-p\cdot u) }{p\cdot u}\np 
\stp
Following the convention used in \cite{Dusling:2009df}, 
the canonical relaxation time is proportional to $(-\pu)^{1-\alpha}$, where $\alpha$
lies generally between the quadratic ansatz limit($\alpha=0$) and linear 
ansatz limit ($\alpha=1$). We have
\st
\t C_{\p}=c_0 (-\pu)^{-\alpha},
\stp 
with $c_0$ to be fixed by shear viscosity (see below). 
The  parameters $u^{*\mu}(x)$ and $T^*(x)$ 
which appear in the \Eq{boltzmann} are equal at leading order to 
the temperature and flow velocities ($T$ and $u^\mu$). They satisfy the 
Landau matching conditions, i.e. ,
\st
\label{matching}
\left(\nu\int\frac{d^3p}{(2\pi)^3E_\p}p^\mu p^\nu
\delta f(u^*, T^*)\right)u^*_\mu=0,
\qquad
\left(\nu\int\frac{d^3p}{(2\pi)^3E_\p}p^\mu p^\nu
n_\p(u^*, T^*)\right)u^*_\mu=eu^{*\nu}.
\stp
So there exist the expansion
\begin{align}
  T^{*}(X) \equiv& T(X)  +  \delta T^* \nc\\
  u^{*\mu}(X) \equiv & u^\mu(X) + \delta u^{*\mu}\nc
\end{align}
with $\delta T^{*}=\sum_{n=1} \delta T_{(n)}$ and 
$\delta u^{*\mu}=\sum_{n=1} 
\delta u^\mu_{(n)}$ indicating corrections from all higher order gradients. 
Expanding $n_{p}^*$ (with an obvious notation) we find correspondingly,
\st
 n_\p^* \equiv n_\p +  \delta n_{p}^*  \quad\mbox{and}\quad 
\delta n_{\p}^* = n_\p'\left[ \frac{p\cdot \delta u_{*}}{T} - E_\p \frac{\delta T_{*} }{T^2} \right],
\stp
where here and in the following a prime stands for the derivative on $\pu$.
$\delta n_\p^*$ with higher order gradients is important for the 
determination of viscous correction $\delta f$, as will be shown in the next section.  
\Eq{matching}, together with the correspondence between kinetic theory and hydrodynamics,
\st
\label{corres}
\nu\int\frac{d^3p}{(2\pi)^3E_\p}p^\mu p^\nu
\delta f(u^*, T^*)=\pi^{\mu\nu}\nc
\stp
constrains the solution of Boltzmann equation $f(x,\p)$.

\section{Determination of $\delta f$}
\label{df_solution}

We substitute the expansion \Eq{Expansion} 
into the kinetic equation \Eq{boltzmann} and equate orders. 
In doing so we use the hydrodynamic equations of motion 
to write time derivatives of $T(x)$ and $u(x)$ in terms of 
spatial gradients of these fields, see \Eq{eom}. Indeed, with the help of these equations of 
motion, up to first order in gradient, 
\st
\label{lhsboltz}
 p^{\mu} \partial_\mu n_\p = 
  +  n_\p' \frac{ p^{\mu} p^{\nu}} {2 T } \sigma_{\mu\nu}
+ O(\nabla^2)\np
\stp
The terms linearly proportional to the $\sigma_{\mu\nu}$ 
 is ultimately responsible for shear viscosity.

Collecting all the terms of first order in gradient for $\delta f$ we obtain the
preliminary first order solution 
\begin{align}
\label{df1_1}
\delta f_{(1) }  =&  -\t C_\p  \; n_\p'  \; \frac{p^{\mu}p^{\nu}  \sigma_{\mu\nu}  }{2 T^3}+\delta n_{\p(1)}^*. 
\end{align}
But \Eq{df1_1} is not completed until the constant coefficient $c_0$, 
the form of $\delta u^\mu_{(1)}$ and $\delta T_{(1)}$ are 
determined through constraint conditions \Eq{matching} and \Eq{corres}. Obviously there exist the trivial solution
$\delta u^\mu_{(1)}=0$ and $\delta T_{(1)}=0$ to \Eq{matching}, and then
from \Eq{corres} we find,
\st
\label{c0}
\eta=\frac{T^3}{15}B_1.
\stp
$B_1$ is a dimensionless constant from the integral, 
\st
B_1(\alpha)=\nu\int\frac{d^3\t p}{(2\pi)^3\t E_\p}(\pu)^4 [n_\p'\t C_\p].
\stp
It only depends on the underlying microscopic 
dynamics in terms of $\alpha$. We will encounter three more similar constants,
\begin{align}
B_2(\alpha)&=\nu\int\frac{d^3\t p}{(2\pi)^3\t E_\p}(\pu)^6 [n_\p'\t C_\p]'
\t C_p\nc\\
B_3(\alpha)&=-\nu\int\frac{d^3\t p}{(2\pi)^3\t E_\p}(\pu)^5 [n_\p'\t C_\p]
\t C_p\nc\\
B_4&=-\nu\int\frac{d^3\t p}{(2\pi)^3\t E_\p}(\pu)^3 [n_\p']\np
\end{align}
$c_0$ is then the solution to \Eq{c0}, and it is interesting to know that for a massless
gas and in the quadratic ansatz limit, $c_0=\eta/s$. 

We extend the derivation to second order $\delta f_{(2)}$, then
\begin{align}
\label{df2_1}
\delta f_{(2) }  = 
 & -\t C_\p \; n_\p'  \; 
\frac{\eta}{s}  \;  \frac{1}{T^4}
 \Big[  p\cdot u \, p_{\mu} \left(2\llangle\sigma^{\mu\nu}\nabla_\nu lnT\rrangle
+\llangle\Delta^\mu_\alpha\nabla_\beta
\sigma^{\alpha\beta}\rrangle\right)  
   + \frac{(p\cdot u)^2}{4(d-1)} \sigma^2  \Big]\nonumber \\&
-\frac{\t C_\p}{T^2}  \; p^{\mu} \partial_{\mu} \delta f_{(1)}+\delta n_{\p(2)}^*\np 
\end{align}
Clearly there are three sources of contributions, those generated from $\delta f_{(1)}$
in Boltzmann equation, from second order corrections to flow velocity and temperature and
from second order hydrodynamics equations of motion. The dominant term is from $p^\mu\partial_\mu
\delta f_{(1)}$, since in Boltzmann equation the derivative gives rise to one higher
order dependence on $p_T$ simultaneously. To write $\delta f_{(2)}$ 
explicitly we decompose the resulting tensors into 
irreducible tensors of the rotation group in the LRF ; 
\cite{Yan:2012b} provides a few more details. Straightforward then from (\ref{df2_1}), 
with somewhat tedious algebra, we obtain,  
\begin{align}
\label{df2_2}
\delta f_{(2)}
=\;&\frac{[n'\t C_\p]'\t C_\p}{4T^2}\t p^\mu\t p^\nu \t p^\alpha
\t p^\beta\llangle\sigma_{\mu\nu}\sigma_{\alpha\beta}\rrangle-
\frac{[n'\t C_\p]\t C_\p}{2T^2}\t p^\mu\t p^\nu\t p^\alpha
\left[3\llangle\sigma_{\mu\nu}\nabla_{\alpha }
lnT\rrangle+\llangle\nabla_{\alpha}\sigma_{\mu\nu}\rrangle\right]\nonumber\\
&+\t p^\mu\t p^\nu\llangle\sigma_{\mu}^\lambda\sigma_{\nu\lambda}\rrangle
\left[
\frac{[n'\t C_\p]'\t C_\p}{7T^2}(\pu)^2+\frac{[n'\t C_\p]\t C_\p}
{2T^2}\pu\right]\nonumber\\
&+\t p^\nu\t p^\nu\llangle\sigma_{\mu\lambda}\Omega^\lambda_{\nu}\rrangle
\frac{[n'\t C_\p]\t C_\p}{T^2}\pu
-\t p^\mu\t p^\nu\left[\llangle 
D\sigma_{\mu\nu}\rrangle +\frac{\sigma_{\mu\nu}}
{3}\nabla\cdot u\right]\frac{[n'\t C_\p]\t C_\p}{2T^2}\pu\nonumber\\
&+\t p_\mu\pu\left[\frac{\eta}{s}\frac{n'\t C_\p}{T^2}
+\frac{[n'\t C_\p]\t C_\p}{5T^2}\pu\right]
\left[2\llangle\sigma^{\mu\nu}\nabla_\nu lnT\rrangle
+\llangle\Delta^\mu_\alpha\nabla_\beta
\sigma^{\alpha\beta}\rrangle\right]\nonumber\\
&+\sigma^2\left[-\frac{\eta}{s}\frac{n'\t C_\p}{T^2}\frac{(\pu)^2}
{12}+\frac{[n'\t C_\p]'\t C_\p}{4T^2}\frac{2(\pu)^4}{15}
+\frac{[n'\t C_\p]\t C_\p}{2T^2}\frac{(\pu)^3}{3}\right]\nonumber\\
&+\delta n_{\p(2)}^* \np
\end{align}
The first line in (\ref{df2_2}) does not contribute to $\pi^{\mu\nu}$, since
all tensor structure from momentum integral are orthogonal to these terms
in brackets. 
Besides the already fixed parameter $c_0$, we still need to determine
second order corrections to flow velocity and temperature, by solving 
Landau-Lifshitz matching condition. And we find that these following solutions 
satisfy all the constraint conditions, up to second order in gradient, 
\begin{align}
\label{ut_2}
\delta u^\mu_{(2)}&=-\frac{3}{B_4T^2}
\left[-\frac{\eta}{3s}B_1+\frac{B_3}{12}\right]
\left[2\llangle\sigma^{\mu\nu}\nabla_\nu lnT\rrangle
+\llangle\Delta^\mu_\alpha\nabla_\beta
\sigma^{\alpha\beta}\rrangle\right]
\\
\delta T_{(2)}&=-\frac{\sigma^2}{B_4T}\left[-\frac{\eta}{s}\frac{B_1}{12}
+\frac{B_2}{30}-\frac{B_3}{6}\right]\np
\end{align}   
As a by-product of the constraint condition \Eq{corres}, we also recognize that
\begin{equation}
\label{trans_coe}
\eta\tau_\pi=\frac{T^2B_3}{15},\quad
\lambda_1=\eta\tau_\pi\left[\frac{2B_2}{7B_3}+1\right],\quad
\lambda_2=-2\eta\tau_\pi,\quad
\lambda_3=0.
\end{equation} 
These relations are in the consistent range with those discussed in \cite{York:2008rr}. 
In particular, all these non-trivial second order transport coefficients are 
non-free parameters in our formalism, as long as $\eta/s$ is fixed. 
This reduces the number of free parameters in hydrodynamical calculations, and
thus quite significant for the recent study on the extraction of $\eta/s$\cite{Shen:2011zc}.

\section{Viscous Hydrodynamic Simulation and Elliptic Flow}
\label{hydro_res}

The simulation of hydrodynamics in code needs solving hydrodynamical equations
of motion. For viscous hydrodynamics, for the convenience of computation, the
algorithm is introduced by approximation. In BRSSS formalism, up to second
order in gradients, \Eq{stress_2} can be rewritten as
\st
\label{pi_eom}
\pi^{\mu\nu}=-\eta\sigma^{\mu\nu}-\tau_\pi\left[\bra D\pi^{\mu\nu}\ket
+\frac{4}{3}\pi^{\mu\nu}\nabla\cdot u
\right]-\frac{\lambda_1}{\eta}\bra\pi_{\lambda}^{\;\mu}\sigma^{\lambda\nu}\ket
-\frac{\lambda_2}{\eta}\bra\pi_\lambda^{\;\mu}\Omega^{\lambda\nu}\ket
+O(\nabla^3),
\stp  
such that a linearized equation for $\pi^{\mu\nu}$, which is approximately
correct if all higher order gradients are neglected, is obtained. Then 
together with general hydrodynamical equations of motion, $\partial_\mu T^
{\mu\nu}$=0, and appropriately selected equation of state, we have a complete
set of equations for the unknown variables, $(\pi^{\mu\nu}, e, u^\mu)$. 
There are two options for the equation of state in our calculations. For theoretical
interest we can consider conformal equation of state(CEOS). 
Although CEOS does not have the accurate description for real medium expansion in heavy ion collisions,
it has the conformal symmetry as assumed in BRSSS hydrodynamics. The previously used 
lattice equation of state(LEOS) by Romatschke and Luzum\cite{Luzum:2008cw} is also tested. 
The details of the algorithm can be found in \cite{Yan:2012b}. For the initial state,
smooth initial condition is considered. As indicated by several groups\cite{Qiu:2011iv,Gardim:2011xv}, 
event fluctuations in the calculations of $v_2$ can be safely
ignored. A constant temperature freeze out scheme is taken, with $T_{fo}=150$ 
MeV.

\begin{figure}[h]
\begin{center}
\includegraphics[width=32pc]{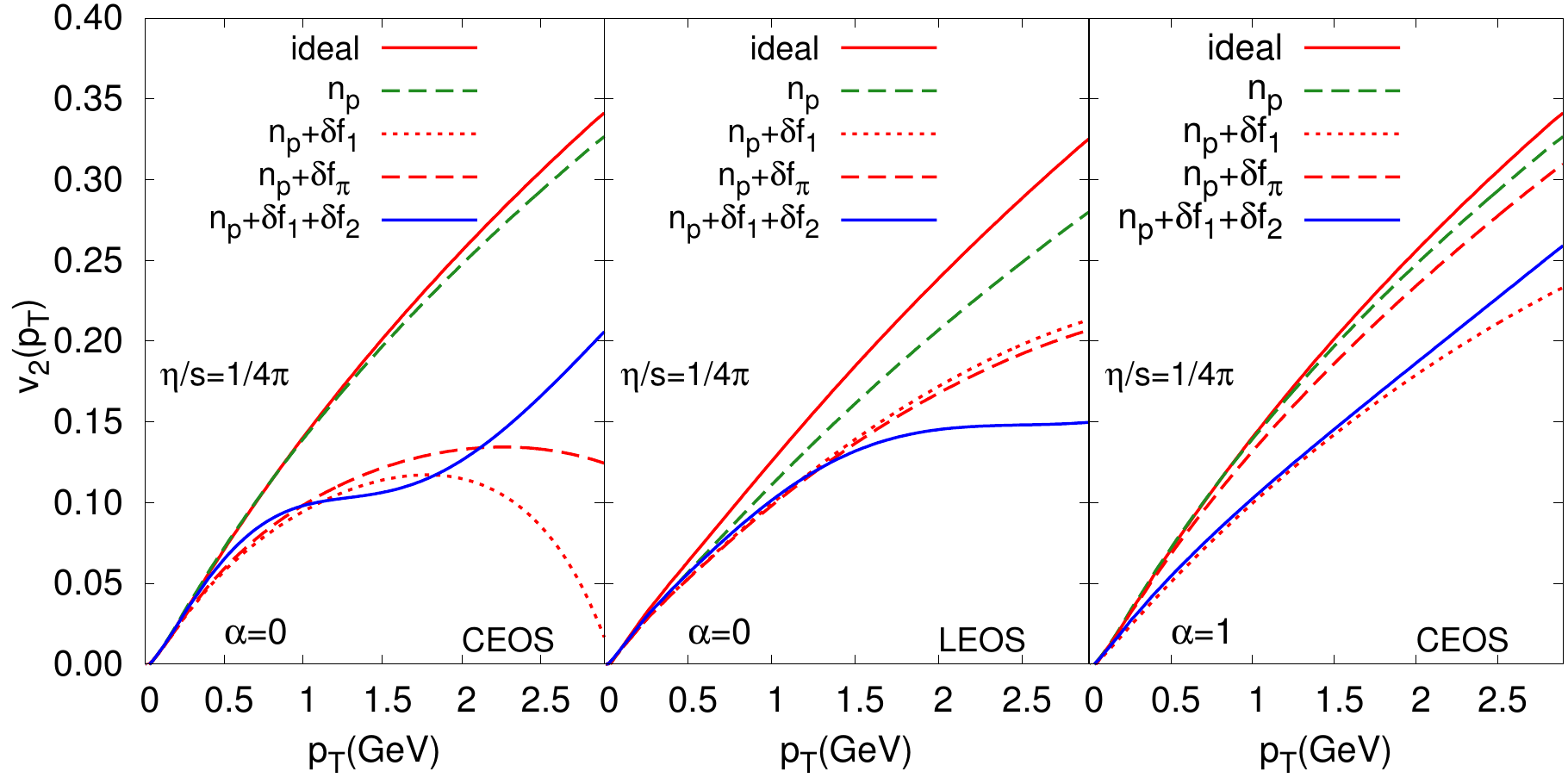}
\caption{\label{label} Differential $v_2$ results from various viscous hydrodynamical
calculations, $b=6.8$ fm.}
\label{v2}
\end{center}
\end{figure}
\begin{figure}[h]
\begin{center}
\includegraphics[width=32pc]{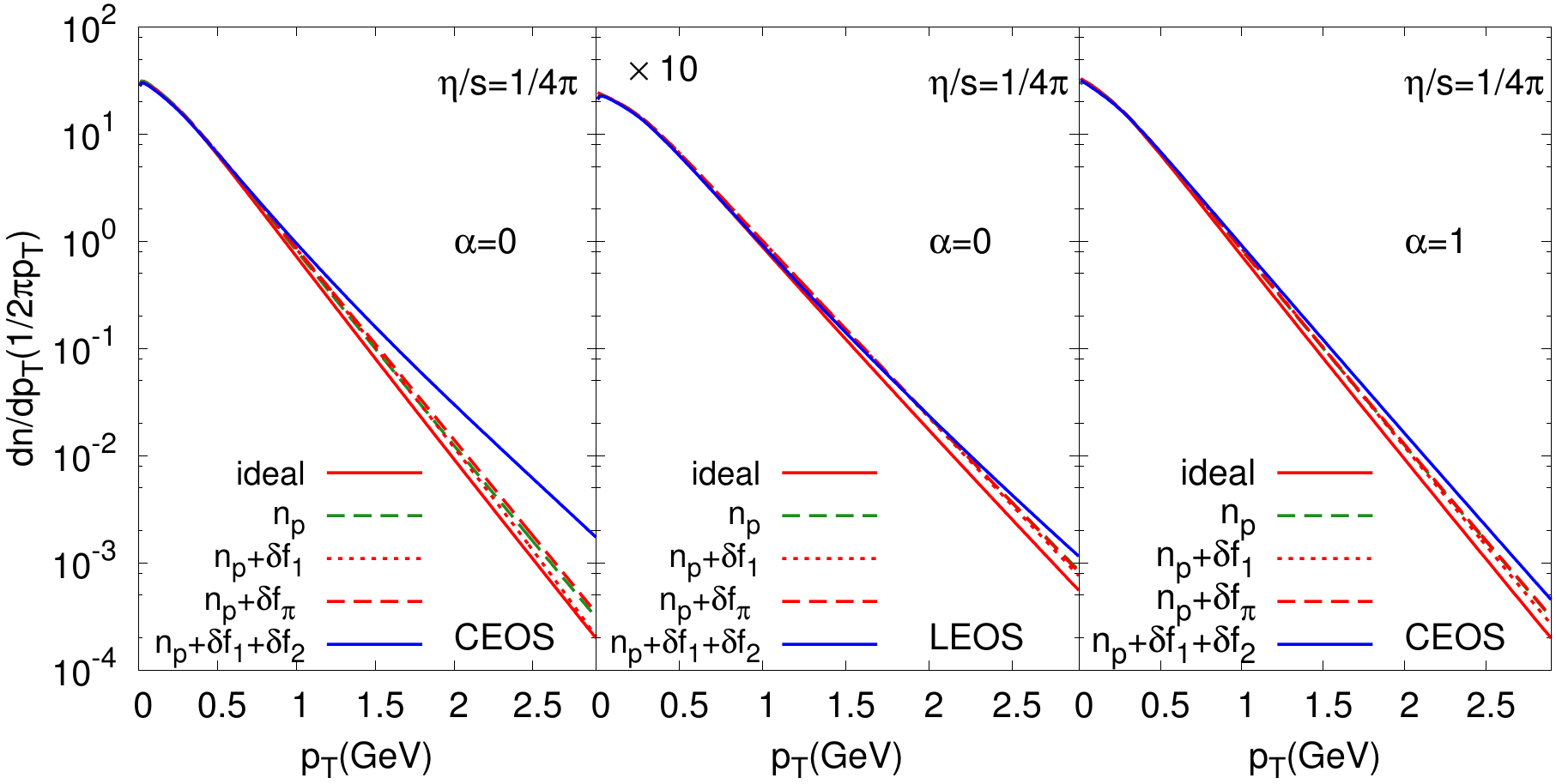}
\caption{\label{label}Spectrum from various viscous hydrodynamical
calculations, $b=6.8$ fm. Note that a factor of 10 has been multiplied in the 
case of LEOS to make it on the same scale.}
\label{dn}
\end{center}
\end{figure}

The effects of including consistent $\delta f$ in freeze out procedure can be
best seen in the results of elliptic flow. In \Fig{v2} the results of a 
series of calculations are shown. Compare to ideal hydrodynamics, viscous 
hydrodynamics indeed damps the generated $v_2$, and
a major fraction of the damping is from $\delta f$. This is reflected in 
the difference between the calculated $v_2$ with equilibrium distribution function 
$n_p$ and that with $n_p+\delta f$'s. The complete and consistent calculation with respect to
BRSSS hydrodynamics needs also $\delta f_{(2)}$ correction. 
However, in the quadratic ansatz limit and conformal equation of state the 
corresponding $v_2$(left panel in \Fig{v2}) 
quickly becomes unreliable when $p_T$ goes up to $1.5$ GeV. This 
is due to the dramatic increase of viscous corrections at large $p_T$ region, and
has already been discussed in the first order case with $\delta f_\pi$\cite{Song:2007ux}. 
But now that
$\delta f_{(2)}$ has higher $p_T$($p_T^4$) dependence, this effect is much stronger.
In another way, we can estimate the magnitude of viscous corrections by 
calculating the spectrum. Then gradient expansion formalism fails when
$\delta f$ contributes to spectrum no longer perturbatively. As seen in the left panel in
\Fig{dn}, we find similarly for CEOS and quadratic ansatz limit, $p_T>1.5$GeV 
region is beyond the feasibility of our formalism. 

Changing equation of state or the $\alpha$ dependence, however, can help solving the
problem to some extent. As seen in the two other panels in \Fig{v2} and \Fig{dn}, viscous corrections
are smaller. On one hand these changes extend our reliable calculation to higher 
$p_T$ region. On the other hand, realistic medium in heavy ion collisions 
has equation of state more close to LEOS, and also quadratic ansatz limit does not
necessarily reflect the real microscopic dynamics. So as expected, second
order corrections in $\delta f$ lead to extra corrections to calculated observables.

\section{Discussions and Future Work}
\label{con}

We have presented a formalism for the derivation of viscous corrections to
phase space distribution function at freeze out. With respect to BRSSS 
hydrodynamics, from this formalism we have obtained the consistent form of 
$\delta f_{(2)}$, and checked its impact on hydrodynamical simulations. $\delta f_{(2)}$
affects the results as corrections, and the corrections increase with $p_T$.
But for LEOS or $\alpha=1$, the corresponding corrections are as small as 
perturbations. 
This is expected from the gradient expansion of the formalism we are following. 
At last, this dependence on $\alpha$ verifies the claim in \cite{Dusling:2009df}
that quadratic ansatz may be questionable in real calculations. 

One important property of higher order viscous corrections in $f(x,\p)$, is 
its higher $p_T$ dependence. This plays a crucial role in the study of 
anisotropic flow for two aspects. 1. This makes the flow results more sensitive
to $\eta/s$. 2. This makes higher order anisotropic flow $v_n$ more sensitive.  
Both of these aspects have positive effects on the extraction of $\eta/s$ from
heavy ion collisions. And this will be the main subject of our future work.

\section*{Acknowledgements}
This work is supported in part by the Sloan Foundation and by the Department of Energy 
through the Outstand Junior Investigator programm DE-FG-02-08ER4154.

\section*{References}
\bibliography{iopart-num}

\providecommand{\newblock}{}
\begin{thebibliography}{10}
\expandafter\ifx\csname url\endcsname\relax
  \def\url#1{{\tt #1}}\fi
\expandafter\ifx\csname urlprefix\endcsname\relax\def\urlprefix{URL }\fi
\providecommand{\eprint}[2][]{\url{#2}}

\bibitem{Alver:2010gr}
Alver B and Roland G 2010 {\em Phys.Rev.\/} {\bf C81} 054905 (\textit{Preprint}
  \eprint{1003.0194})

\bibitem{Schenke:2010rr}
Schenke B, Jeon S and Gale C 2011 {\em Phys.Rev.Lett.\/} {\bf 106} 042301
  (\textit{Preprint} \eprint{1009.3244})

\bibitem{Qiu:2011iv}
Qiu Z and Heinz U~W 2011 {\em Phys.Rev.\/} {\bf C84} 024911 (\textit{Preprint}
  \eprint{1104.0650})

\bibitem{Teaney:2010vd}
Teaney D and Yan L 2011 {\em Phys.Rev.\/} {\bf C83} 064904 (\textit{Preprint}
  \eprint{1010.1876})

\bibitem{Alver:2010dn}
Alver B~H, Gombeaud C, Luzum M and Ollitrault J~Y 2010 {\em Phys.Rev.\/} {\bf
  C82} 034913 (\textit{Preprint} \eprint{1007.5469})

\bibitem{Teaney:2003kp}
Teaney D 2003 {\em Phys.Rev.\/} {\bf C68} 034913 (\textit{Preprint}
  \eprint{nucl-th/0301099})

\bibitem{Danielewicz:1984ww}
Danielewicz P and Gyulassy M 1985 {\em Phys.Rev.\/} {\bf D31} 53--62

\bibitem{Dusling:2007gi}
Dusling K and Teaney D 2008 {\em Phys.Rev.\/} {\bf C77} 034905
  (\textit{Preprint} \eprint{0710.5932})

\bibitem{Luzum:2008cw}
Luzum M and Romatschke P 2008 {\em Phys.Rev.\/} {\bf C78} 034915
  (\textit{Preprint} \eprint{0804.4015})

\bibitem{Song:2007ux}
Song H and Heinz U~W 2008 {\em Phys.Rev.\/} {\bf C77} 064901 (\textit{Preprint}
  \eprint{0712.3715})

\bibitem{Romatschke:2009im}
Romatschke P 2010 {\em Int.J.Mod.Phys.\/} {\bf E19} 1--53 (\textit{Preprint}
  \eprint{0902.3663})

\bibitem{Baier:2007ix}
Baier R, Romatschke P, Son D~T, Starinets A~O and Stephanov M~A 2008 {\em
  JHEP\/} {\bf 0804} 100 (\textit{Preprint} \eprint{0712.2451})

\bibitem{Yan:2012b}
Teaney D and Yan L   , In progress

\bibitem{DeGroot:1980dk}
De~Groot S, Van~Leeuwen WA e and Van~Weert C 1980

\bibitem{Dusling:2009df}
Dusling K, Moore G~D and Teaney D 2010 {\em Phys.Rev.\/} {\bf C81} 034907
  (\textit{Preprint} \eprint{0909.0754})

\bibitem{Policastro:2001yc}
Policastro G, Son D and Starinets A 2001 {\em Phys.Rev.Lett.\/} {\bf 87} 081601
  (\textit{Preprint} \eprint{hep-th/0104066})

\bibitem{York:2008rr}
York M~A and Moore G~D 2009 {\em Phys.Rev.\/} {\bf D79} 054011
  (\textit{Preprint} \eprint{0811.0729})

\bibitem{Shen:2011zc}
Shen C, Bass S~A, Hirano T, Huovinen P, Qiu Z {\em et~al.\/} 2011 {\em
  J.Phys.G\/} {\bf G38} 124045 (\textit{Preprint} \eprint{1106.6350})

\bibitem{Gardim:2011xv}
Gardim F~G, Grassi F, Luzum M and Ollitrault J~Y 2012 {\em Phys.Rev.\/} {\bf
  C85} 024908 (\textit{Preprint} \eprint{1111.6538})

\end{thebibliography}

\end{document}